\documentclass[9pt,twocolumn,twoside]{pnas-new}

\templatetype{pnasresearcharticle} 

\usepackage[normalem]{ulem}
\usepackage{pdfpages}

\title{Propagation of pop~ups in kirigami shells}

\author[a,b,1]{Ahmad Rafsanjani}
\author[a,c,1]{Lishuai Jin} 
\author[a]{Bolei Deng} 
\author[a,d,e,2]{Katia Bertoldi}

\affil[a]{John A. Paulson School of Engineering and Applied Sciences, Harvard University, Cambridge, MA 02138, USA}
\affil[b]{Department of Materials, ETH Z\"urich, 8093 Z\"urich, Switzerland}
\affil[c]{Department of Mechanics, Tianjin University, Tianjin 300072, China}
\affil[d]{Kavli Institute, Harvard University, Cambridge, MA 02138, USA}
\affil[e]{Wyss Institute for Biologically Inspired Engineering, Cambridge, MA 02138, USA}

\leadauthor{Rafsanjani} 

\significancestatement{Kirigami - the Japanese art of cutting paper - has become emergent tool to realize  highly stretchable devices  and morphable structures. While kirigami structures are fabricated by simply perforating an array of cuts into a thin sheet,  the applied deformation and associated instabilities can be exploited to transform them into complex 3D morphologies. However, to date, such reconfiguration always happen simultaneously through the system. By borrowing ideas from phase-transforming materials, we combine cuts and curvature to realize kirigami structures in which the deformation-induced shape reconfiguration initially nucleates near an imperfection and then, under specific conditions, spreads through the system.  We envision that such control of the shape-transformation could be used to design the next generation of responsive surfaces and smart skins.}

\authorcontributions{A.R., L.J., and K.B. designed research; A.R., L.J., and B.D. performed research; A.R., L.J., and B.D. analyzed data; B.D. developed the theoretical model; and A.R., L.J., B.D., and K.B. wrote the paper.}
\authordeclaration{Please declare any conflict of interest here.}
\equalauthors{\textsuperscript{1}A.R. and L.J. contributed equally to this work.}
\correspondingauthor{\textsuperscript{2}To whom correspondence should be addressed. E-mail: bertoldi@seas.harvard.edu}

\keywords{kirigami $|$ buckling $|$ propagative instability $|$ metamaterials $|$ phase transition} 

\begin{abstract}
Kirigami-inspired metamaterials are attracting increasing interest because of their ability to achieve  extremely large strains and shape changes via out-of-plane buckling. 
While in  flat kirigami sheets the ligaments buckle simultaneously as Euler columns  leading to a continuous phase transition, here we demonstrate that kirigami shells can also support discontinuous phase transitions. Specifically, we show via a combination of experiments, numerical simulations and theoretical analysis that in cylindrical kirigami shells the snapping-induced curvature inversion of the initially bent ligaments results in a pop-up process that first localizes near an imperfection and then, as the deformation is increased, progressively spreads through the structure.
Notably, we find that the width of the transition zone as well as the stress at which propagation of the instability is triggered can be controlled by carefully selecting the geometry of the cuts and the curvature of the shell. Our study significantly expands the ability of existing kirigami metamaterials and opens avenues for the design of the next generation of responsive surfaces, as demonstrated by the design of a smart skin that significantly enhance the crawling efficiency of a simple linear actuator.
\end{abstract}

\dates{This manuscript was compiled on \today}
\doi{\url{https://doi.org/10.1073/pnas.1817763116}}
\begin{document}

\maketitle
\thispagestyle{firststyle}
\ifthenelse{\boolean{shortarticle}}{\ifthenelse{\boolean{singlecolumn}}{\abscontentformatted}{\abscontent}}{}

\dropcap{K}irigami -- the Japanese art of cutting  paper -- has recently inspired the design of  highly stretchable~\cite{Cho2014, Zhang2015, Shyu2015, Blees2015, Isobe2016, Dias2017, Rafsanjani2018, Hwang2018} and morphable~\cite{Sussman2015, Neville2016, Konakovic2016, Castle2016, Rafsanjani2017, Zhao2017, Yang2018, Liu2018, Yang2018b} mechanical metamaterials that  can be easily realized by embedding an array of cuts into a  thin sheet. An attractive feature of these systems is that they are manufactured flat and then exploit elastic instabilities to transform into complex three-dimensional configurations  ~\cite{Zhang2015, Blees2015, Shyu2015, Isobe2016, Rafsanjani2017, Dias2017}.  Remarkably,  the morphology of such buckling-induced 3D patterns can be tuned  by varying the arrangement and geometry of the cuts~\cite{Blees2015, Shyu2015, Zhang2015} as well as the loading direction~\cite{Rafsanjani2017}.   However, in all kirigami systems proposed to date the buckling-induced pop-up process occurs concurrently through the entire system, resulting in a simultaneous shape transformation. 

The coexistence of two phases has been observed both at the microscopic and macroscopic scale in a variety of systems, including  phase transforming materials~\cite{Ericksen1975, James1979, Abeyaratne1991, Shaw1997, Feng2006}, dielectric elastomers~\cite{Zhao2007, Zhou2008} and thin-walled elastic tubes~\cite{Chater1984, Kyriakides1993} (see Movie S1). While these systems are very different in nature from each other, they all share a non-convex free energy function that for specific conditions has two minima of equal height. When such situation is reached, the homogeneous deformation becomes
unstable, and  a mixture of two states emerges.  The new phase initially nucleates near a local imperfection and then, under prevailing conditions, propagates through the entire system~\cite{Shaw1997, Chater1984, Kyriakides1993, Zhou2008}.

Here, we demonstrate via a combination of experiments and numerical/theoretical analyses that kirigami structures can also support the coexistence of two phases, the buckled and unbuckled one. Specifically, we show that in thin cylindrical kirigami shells subjected to tensile loading the buckling-induced pop-up process initially localizes near an imperfection and then, as the deformation is increased,  progressively spreads through the cylinder at constant stress. 
We find that the curvature of the cylinder is the essential ingredient to observe this phenomenon, as it completely change the deformation mechanism of the hinges. In kirigami sheets the initially flat hinges  buckle out-of-plane, leading to a monotonic stress-strain relationship for the unit cell. By contrast, in  kirigami shells the initially bent ligaments  snap to their second stable configuration, resulting in a non-monotonic stress-strain curve  typical of phase-transforming materials~\cite{Ericksen1975,James1979, Abeyaratne1991, Shaw1997, Feng2006}.

\begin{figure}[!t]
\centering
\includegraphics [width=\linewidth]{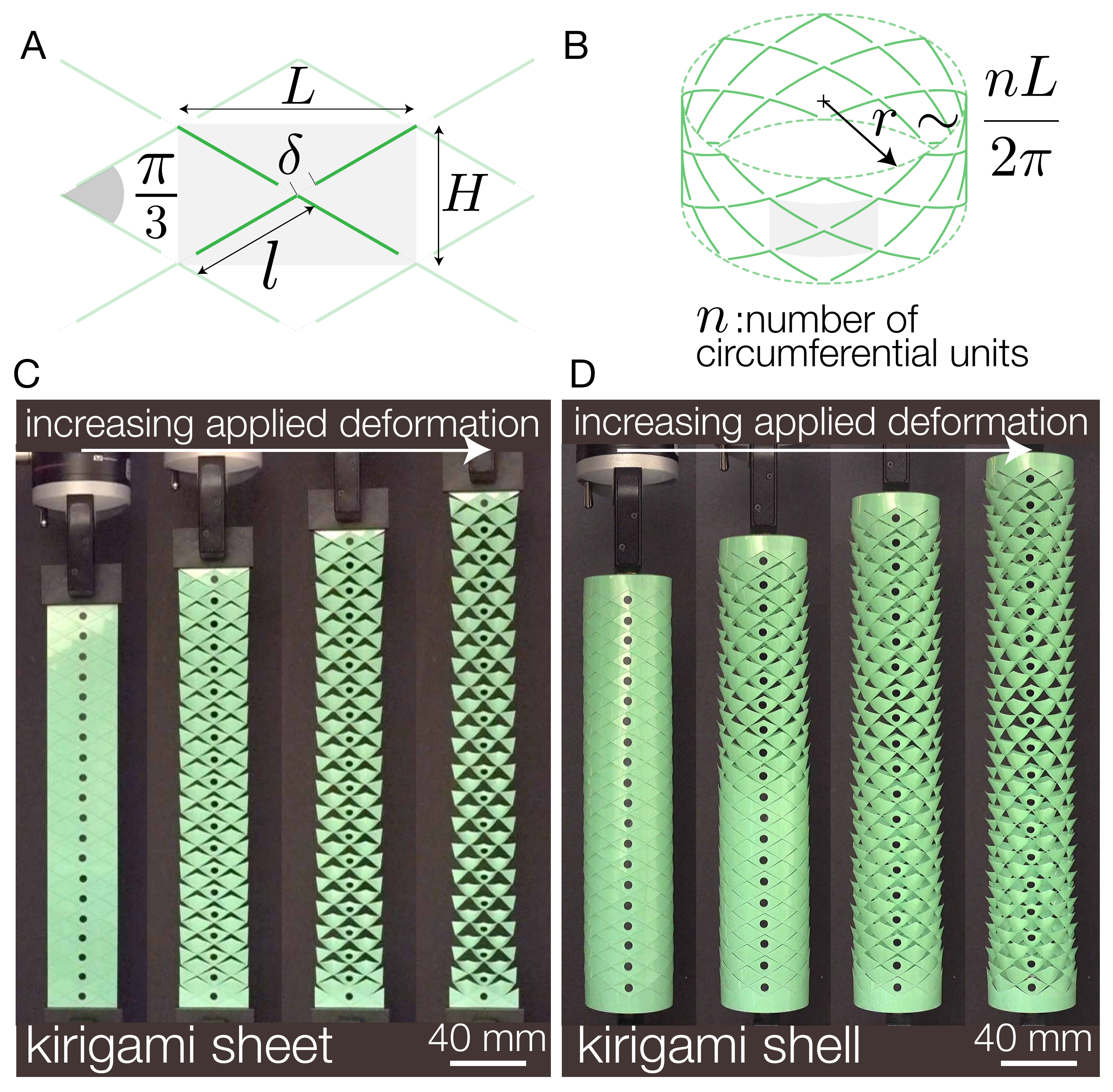}
\caption{(\textit{A}) Schematic of a kirigami structure that comprise an array of triangular cuts. The unit cell has width $L$ and height $H$ and is highlighted in grey.  (\textit{B}) Schematic of a kirigami cylindrical shell comprising $n=8$ unit cells along the circumference. (\textit{C}) Experimental  images of a kirigami sheet with a triangular pattern characterized by $\delta/l=0.0625$ (with $l=12\:$mm) at different
levels of deformation. The  pop-up process occurs concurrently through the entire system.  (\textit{D}) Experimental  images of a kirigami sheet fabricated by rolling a sheet with a triangular pattern characterized by $\delta/l=0.0625$ (with by $l=12\:$mm and $n=8$ unit cells along the circumference) at different levels of deformation. The pop-up process initiates at the top end of the sample and then spreads towards the bottom one as $\bar\varepsilon$ is increased. The thickness of all kirigami structures is $t=76.2\:\mu$m.}
\label{Fig1}
\end{figure}


\section*{Experiments}
We start by  testing  under uniaxial tension a kirigami flat sheet and a corresponding  cylindrical shell. Both structures are fabricated by laser cutting triangular cuts arranged on a triangular lattice with lattice constants $l=12$ mm and $\gamma=\pi/3$ separated by hinges with width $\delta=l/16$  (resulting in a rectangular unit cell with width $L=2l\cos \pi/6$ and height $H=2l\sin \pi/6$) into polyester plastic sheets (Artus Corporation, NJ) of thickness $t=76.2\:\mu$m with Young's modulus $E=4.33\:$GPa and Poisson's ratio $\nu=0.4$ (Fig.~\ref{Fig1}\textit{A}).
The flat kirigami sheet comprises an array of $2\times20$  cuts, while the cylindrical shell has $8\times20$ cuts and is created by bending an initially flat perforated sheet into a cylinder with radius  $r\sim n L/(2\pi)$ ($n$ denoting the number of unit cell along the circumference of the cylinder - see Fig.~\ref{Fig1}\textit{B}) and gluing the two overlapping  edges with a thin adhesive layer (see SI Appendix, section~1 and Movie~S2 for fabrication details). 

In Figs.~\ref{Fig1}\textit{C} and \textit{D} we show  snapshots of the kirigami sheet and kirigami shell at  different levels of applied deformation.  We find that the response of the two structures is remarkably different (see Movie~S3). In the kirigami sheet at a critical strain  all triangular features simultaneously pop-up, forming   a uniform 3D textured surface that becomes  more accentuated for increasing deformation (Fig.\ref{Fig1}\textit{C}).
By contrast, in the cylindrical kirigami shell the pop-up process initiates  at the top end of the sample and then   spreads  towards the other end as the applied deformation is increased (Fig.\ref{Fig1}\textit{D}). 
Note that the pop-up process in our shells typically starts at one of the ends of the shell, since these act as imperfection. As a matter of fact,  a local reduction in the size of the hinges has to be introduced to cause the propagation to start from a different location (see SI Appendix, Fig.~S7).

Next, to  better characterize the response of our structures, during the tests we  monitor   black circular markers located at the base of the triangular cuts  (Figs.~\ref{Fig1}\textit{C} and \textit{D}) and use their position to determine both the applied strain, $\bar\varepsilon$, and the local strain in longitudinal direction for the $i$-th row of cuts, $\varepsilon_i$, as   
\begin{equation}
\bar\varepsilon=\frac{z_{q}-z_p}{Z_{q}-Z_p}-1,\;\;\;\;\;\;\varepsilon_i=\frac{z_{i+1}-z_i}{Z_{i+1}-Z_i}-1,
\label{eq_epsilonbar}
\end{equation}
where $z_i$ and $Z_i$ denote the position of the $i$-th marker in the deformed and undeformed  configuration, respectively, and we choose $p=3$ and $q=18$ to minimize boundary effects. Moreover, we use a custom laser profilometer and track the deformation of  an horizontal line  passing through the the hinges at different levels of applied deformation.

\begin{figure}[!t]
\centering
\includegraphics[width=\columnwidth]{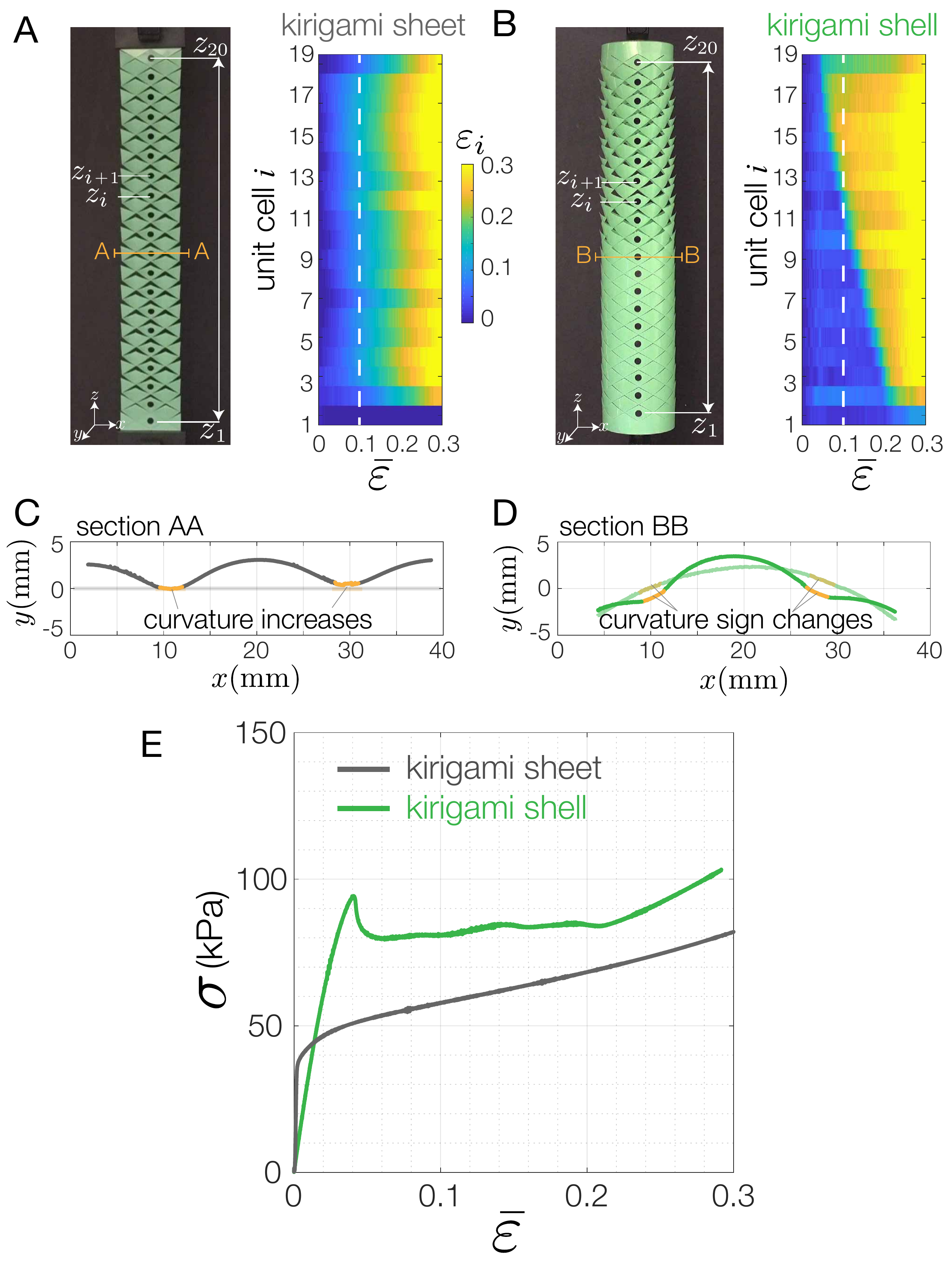}
\caption{ 
(\textit{A})-(\textit{B}) Evolution of the local strain $\varepsilon_i$ as a function of $\bar\varepsilon$ for the (A) kirigami sheet shown in Fig. \ref{Fig1}C.  and (B) the kirigami shell shown in Fig. \ref{Fig1}D. (\textit{C})-(\textit{D}) Projection  in the $xy$-plane of a line passing  through hinges (such line is indicated in orange in A and B) as reconstructed using a custom laser profilometer for (\textit{C}) the kirigami sheet and (\textit{D}) kirigami shell.   The deformation of the hinges is highlighted in orange.  (\textit{E}) Experimental stress-strain curves for the kirigami sheet and kirigami shell considered in Fig.~\ref{Fig1}.  Note that the amplitude of the peak observed for the kirigami shell is correlated to the size of imperfection that triggers the pop-up (see SI Appendix, Fig.~S20).}
\label{Fig2}
\end{figure}

In  Figs.~\ref{Fig2}\textit{A} and \textit{B} we report the evolution of the local strain $\varepsilon_i$ as a function of $\bar\varepsilon$ for the kirigami sheet and shell, respectively. In full agreement with our previous observations, we find that in the kirigami sheet  the local strain increases uniformly  through the structure and is always very close to the applied deformation (i.e.~$\varepsilon_i\sim\bar\varepsilon\,\, \forall\, i$, see Fig.~\ref{Fig2}\textit{A}). 
Differently, the contour map for the  kirigami cylindrical shell shows a non-vertical boundary between popped/open (yellow) and unpopped/closed (blue) regions (Fig.~\ref{Fig2}\textit{B}) - a clear signature of  sequential opening. Furthermore,
the constant slope of such boundary indicates that the pop-up process propagates at constant rate of applied deformation (see SI Appendix, section 2).

To gain more insight into the physics behind the different behavior observed  in the kirigami sheet and kirigami shell, we then investigate the deformation mechanism of their hinges.  By inspecting their 3D-scanned profiles (see Fig.~\ref{Fig2}\textit{C} and \textit{D}), we find that they deform in a very different way. In the kirigami sheet the hinges are initially flat and act as straight beams~\cite{Isobe2016, Rafsanjani2017}; for a critical level of applied deformation they buckle and subsequently bend out-of-plane. By contrast, in the kirigami shell the initially bent hinges behave as bistable arches~\cite{Plaut2015} and snap to their second stable configurations which is characterized by curvature inversion.  This observation is fully consistent with the results of Figs.~\ref{Fig1} and~\ref{Fig2}, since  snapping is always accompanied by a highly non-linear stress-strain response which is typical of phase-transforming materials~\cite{Ericksen1975,James1979, Abeyaratne1991, Shaw1997, Feng2006}. 
As a matter of fact, while elastic structures comprising arrays of beams that buckle under the applied load have been shown to display homogeneous pattern transformations~\cite{Rafsanjani2017, Kang2014}, sequential events are typically observed in systems based on snapping units~\cite{Shan2015, Rafsanjani2015, Restrepo2016}.



Finally, in Fig.~\ref{Fig2}\textit{E} we compare the  stress-strain curves of  the kirigami sheet and the kirigami shell. We find that  the response of the kirigami sheet is typical of buckling-based structures~\cite{Bertoldi2010} and  is  characterized by an initial linear regime (during which all hinges bend in-plane)  followed by a  plateau stress  (caused by the homogeneous buckling-induced  pop-up process).
The cylindrical kirigami shell also exhibits these two regimes, but the transition between them is more abrupt and characterized by a sharp load drop. 
At the peak  a small portion of the  kirigami shell near the top end pops up, causing the unloading of the rest of the structure and a drop in stress. 
Subsequently, when the ligaments of the buckled region start to be stretched and become resistant  to further deformation, the pop-up process spreads sequentially  through the entire structure and the stress reaches a steady state value $\sigma_p=83.5\:$kPa. Finally, once all units are fully popped-up at $\bar\varepsilon\sim 0.22$, the stress starts to rise again because of further stretching of all hinges.

\section*{Modeling}

\begin{figure*}[h]
\centering
\includegraphics[width=\textwidth]{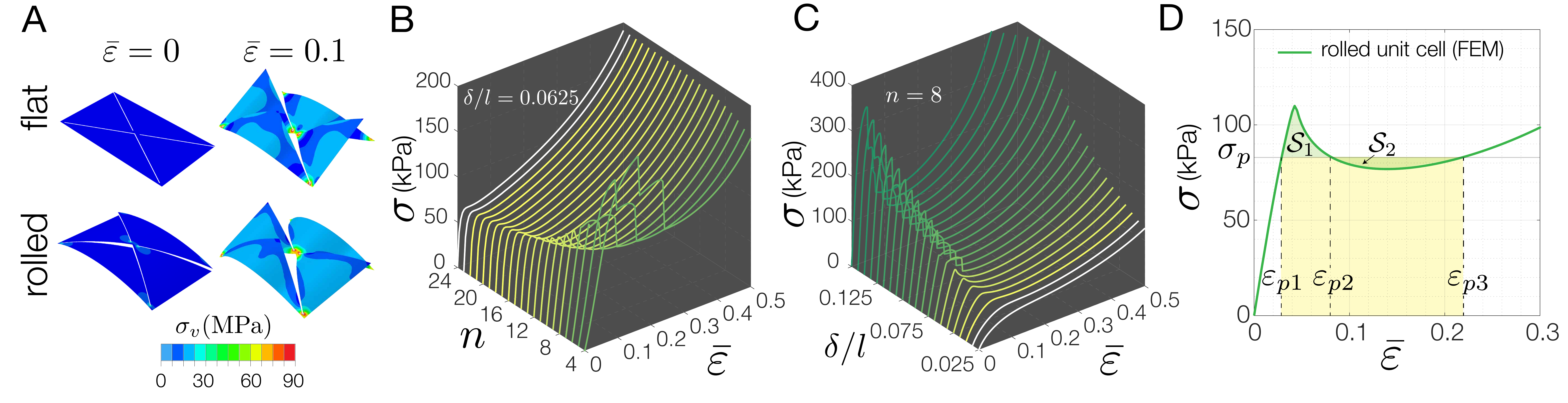}
\caption{\textit{A}) FE snapshots  of a flat and rolled unit cells (characterized by $\delta/l=0.0625$ with $l=12\:$mm)  at  $\bar\varepsilon=0$ and $0.1$.
(\textit{B}) Strain-stress curves of unit cells with $\delta/l=0.0625$ as a function of $n$.
(\textit{C}) Strain-stress curves of unit cells with $n=8$ as a function of $\delta/l$.
(\textit{D}) Stress-strain response of a  unit cells with $\delta/l=0.0625$ and $n=8$. In the plot we highlight the energy barrier $\mathcal{S}_1$, the steady-state propagation stress $\sigma_p$ and the critical strains $\varepsilon_{p1}$, $\varepsilon_{p2}$ and $\varepsilon_{p3}$.
The thickness of all kirigami unit cells is $t=76.2\:\mu$m. }
\label{Fig3}
\end{figure*}

Having understood how the imposed curvature  affects the  deformation mechanism as well as the response of our kirigami structures, we now use a combination of numerical and analytical tools to  quantify this effect.  To begin with, we conduct non-linear Finite Element (FE)  within Abaqus/Standard to investigate the effect of both the ligament width $\delta$ and the curvature $1/r$ (which is proportional to  $1/n$) on the response  of unit cells  subjected to uniaxial tension  (see SI Appendix, section 4).  We find that for $\delta/l\in[0.025,\, 0.15]$ and $n\in[4,\, 24]$ the applied deformation always triggers a buckling instability that induces the pop-up of the triangular features (see Fig.~\ref{Fig3}\textit{A} and Movie~S7). However, the stress-strain response is found to be significantly affected by both  $\delta$ and $n$ (see Figs.~\ref{Fig3}\textit{B} and \textit{C}).  For large values of $n$ (i.e. for small curvatures) all unit cells are characterized by monotonic stress-strain curves (see Fig.~\ref{Fig3}\textit{B}), irrespectively of $\delta/l$.  Differently, below a critical $n$ the stress-strain response becomes non-monotonic, characterized by a peak, a subsequent drop in load and final stiffening.  Further,  we find that by either decreasing $n$ (at constant $\delta/l$ - see Fig.~\ref{Fig3}\textit{B}) or increasing $\delta/l$ (at constant $n$ - see Fig.~\ref{Fig3}\textit{C})   the peak  becomes more accentuated and is eventually  followed by a sharp drop. 

At this point, we want to emphasize that the non-monotonic up-down-up behavior observed for most of our rolled unit cells is typical of elastic structures supporting propagative instabilities~\cite{Chater1984, Kyriakides1993}. Remarkably, it has been shown that the Maxwell construction~\cite{Maxwell1875} can be applied to such stress-strain curves to determine several key parameters that characterize the behavior of our curved kirigami shell~\cite{Chater1984, Kyriakides1993}. Specifically, by equating the area of the two lobes formed by the $\sigma(\varepsilon)$ curve (i.e. by imposing $\mathcal{S}_1=\mathcal{S}_2$- see Fig.~\ref{Fig3}\textit{D}) we can identify ($i$) the propagation stress $\sigma_p$, ($ii$) the energy barrier $\mathcal{S}_1$ and  ($iii$) the critical strains $\varepsilon_{p1}$,  $\varepsilon_{p2}$ and  $\varepsilon_{p3}$ (see Fig.~\ref{Fig3}\textit{D}). 
For $\bar\varepsilon<\varepsilon_{p1}$ the structure deforms homogeneously and all triangular features are unpopped, whereas for $\varepsilon_{p1}<\bar\varepsilon<\varepsilon_{p3}$ the pop-up process initiated at the top end of the sample spreads towards the other end.  

\begin{figure}[!b]
\centering
\includegraphics[width=\columnwidth]{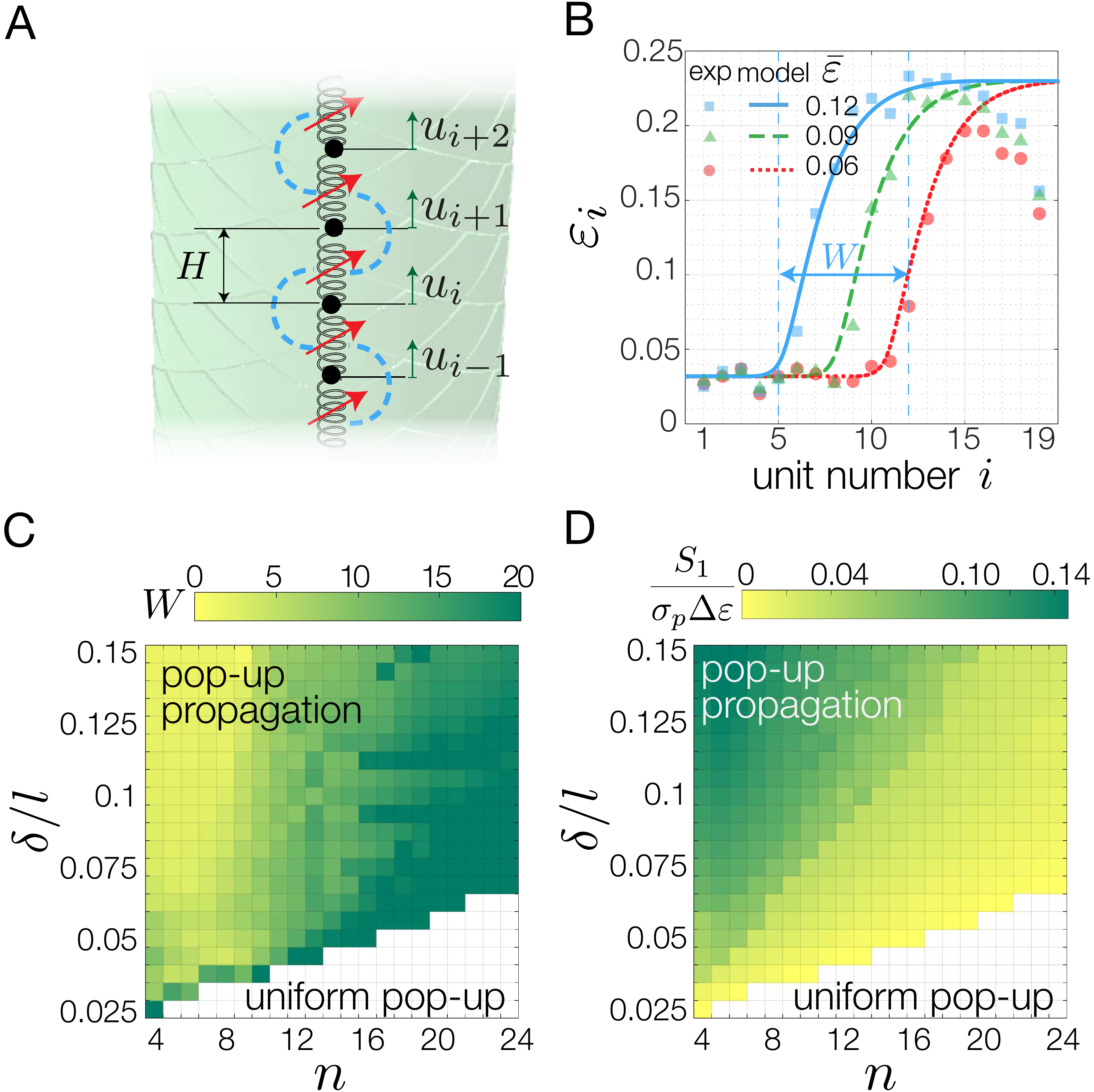}
\caption{
(\textit{A}) Schematic of our  model comprising a 1D array of  non-linear springs. The blue dashed lines indicate the strain gradient interactions that we account for.
(\textit{B}) Comparison between the strain distribution predicted by our model and measured in experiments for the cylindrical kirigami shell shown in Fig.\ref{Fig1}B.
(\textit{C}) Phase diagram of the the width $W$ of the transition zone as predicted by our model.
(\textit{D}) Phase diagram of the energy barrier of the kirigami shells with triangular cut pattern. $\mathcal{S}_1$ (see shaded green area in Fig.~\ref{Fig3}\textit{D}) normalized by the total energy required for phase transition (i.e.~$\sigma_p (\varepsilon_{p3}-\varepsilon_{p1})$) obtained from 441 unit cell FE simulations. The thickness of all kirigami structures is $t=76.2\:\mu$m.
}
\label{Fig4}
\end{figure}

While Maxwell construction enable us to easily determine several parameters, it does not provides any information on the  width and the shape of the transition zone. This motivates the derivation of a more detailed  model  based on  a 1D array of $N$ non-linear springs (see Fig.~\ref{Fig4}A)  in which the  response of the $i$-th element is described as
\begin{equation}
\label{spring}
F_i(u_i,\,u_{i+1}) = nLt\times\sigma\left(\varepsilon_i\right)= nLt\times\sigma\left(\frac{u_{i+1}-u_{i}}{H}\right), 
\end{equation}
where  $\sigma(\varepsilon_i)$ is the non-linear stress-strain response of the unit cell and  $u_{i}=z_i-Z_i$ (see Fig. \ref{Fig4}A).
We then write the strain energy of the system when subjected to a constant force $F_p=\sigma_pnLt$ as
\begin{equation}
\label{energy2}
\begin{split}
\Pi =&U - F_p (u_{N+1} - u_{1})\\
=&\sum_{i=1}^{N} nLt\int^{u_{i+1}-u_i}_{0}\sigma \left(\frac{u_{i+1}-u_{i}}{H}\right) \text{d}(u_{i+1}-u_{i})\\
-&F_p (u_{N+1} - u_{1})+ \sum_{i=2}^{N}  \frac{1}{2} nLt G_{ } \left(u_{i+1} - 2 u_{i} + u_{i-1}\right)^2,
\end{split}
\end{equation}
where the last term (and the coefficient $G$) is introduced to capture the effect of the strain gradient~\cite{Mindlin1964, Polyzos2012,Audoly2016}, which significantly affects the response of our kirigami structures given the strong coupling between their unit cells (see SI Appendix, section 4). It follows that the equilibrium equations are given by 
\begin{equation}
\label{discrete_c}
\begin{split}
\sigma \left(\varepsilon_i\right) &- \sigma \left(\varepsilon_{i-1}\right) - G_{ }\left(\varepsilon_{i+1} - 3 \varepsilon_{i} + 3\varepsilon_{i-1} - \varepsilon_{i-2}\right) = 0,\;\;\;\;\\&\text{for }i=3,\,...,\,N-1
\end{split}
\end{equation}
where $\varepsilon_i$ is defined in Eq.~\textbf{\ref{eq_epsilonbar}}.

Next,  we take the continuum limit of Eqs.~\textbf{\ref{discrete_c}}, retain the nonlinear terms up to third order and integrate it with respect to
$Z$ to obtain 
\begin{equation}
\label{ODE2}
{G H^2}\frac{\text{d}^2\varepsilon}{\text{d}Z^2} =  {\sigma(\varepsilon) - \sigma_p},
\end{equation}
where $Z$ denotes the initial coordinate
along the longitudinal direction, $\varepsilon(Z)$ is a continuous function of $Z$ and we have assumed that at $Z\rightarrow -\infty$ the unit cells are unpopped and subjected to a strain $\varepsilon_{p1}$. Eq.~\textbf{\ref{ODE2}} is the continuum governing equation for our kirigami structures and, given a stress-strain curve of the unit cell $\sigma(\varepsilon)$,  can be numerically solved to obtain the strain distribution $\varepsilon(Z)$ within the structure as a function of the applied strain. To test the relevance of our model, in Fig.~\ref{Fig4}\textit{B} we focus on the cylindrical kirigami shell of Fig.~\ref{Fig1}\textit{B} and compare the evolution of the strain along its axes as predicted by our model and measured in experiments at $\bar\varepsilon=0.06$, 0.09 and 0.12.  Note that the model predictions are obtained by numerically integrating Eq.~\textbf{\ref{ODE2}} with $\sigma(\varepsilon)$ predicted by a FE simulation conducted  on the unit cell, $\sigma_p$ equal to the Maxwell stress and $G=428\:$kPa (see SI Appendix, section 4). Moreover,
since the solution of Eq.~\textbf{\ref{ODE2}} is translational invariant with respect to $Z$, the position of the propagation front $Z_0$  at a given applied strain $\overline\varepsilon$ can be determined from the compatibility condition 
\begin{equation}
\label{spatial_position}
    \bar\varepsilon = \frac{1}{Z_q - Z_p}\int^{Z_q}_{Z_p} \varepsilon(Z) \; \text{d}Z .
\end{equation}
where $p=3$ and $q=18$ (see Eq.~\textbf{\ref{eq_epsilonbar}}). 
We find that our model accurately captures the shape, width and amplitude of the transition zone as well as its position as a function of the  applied strain, confirming the validity of our approach.

\begin{figure}[!b]
\centering
\includegraphics[width=0.9\columnwidth]{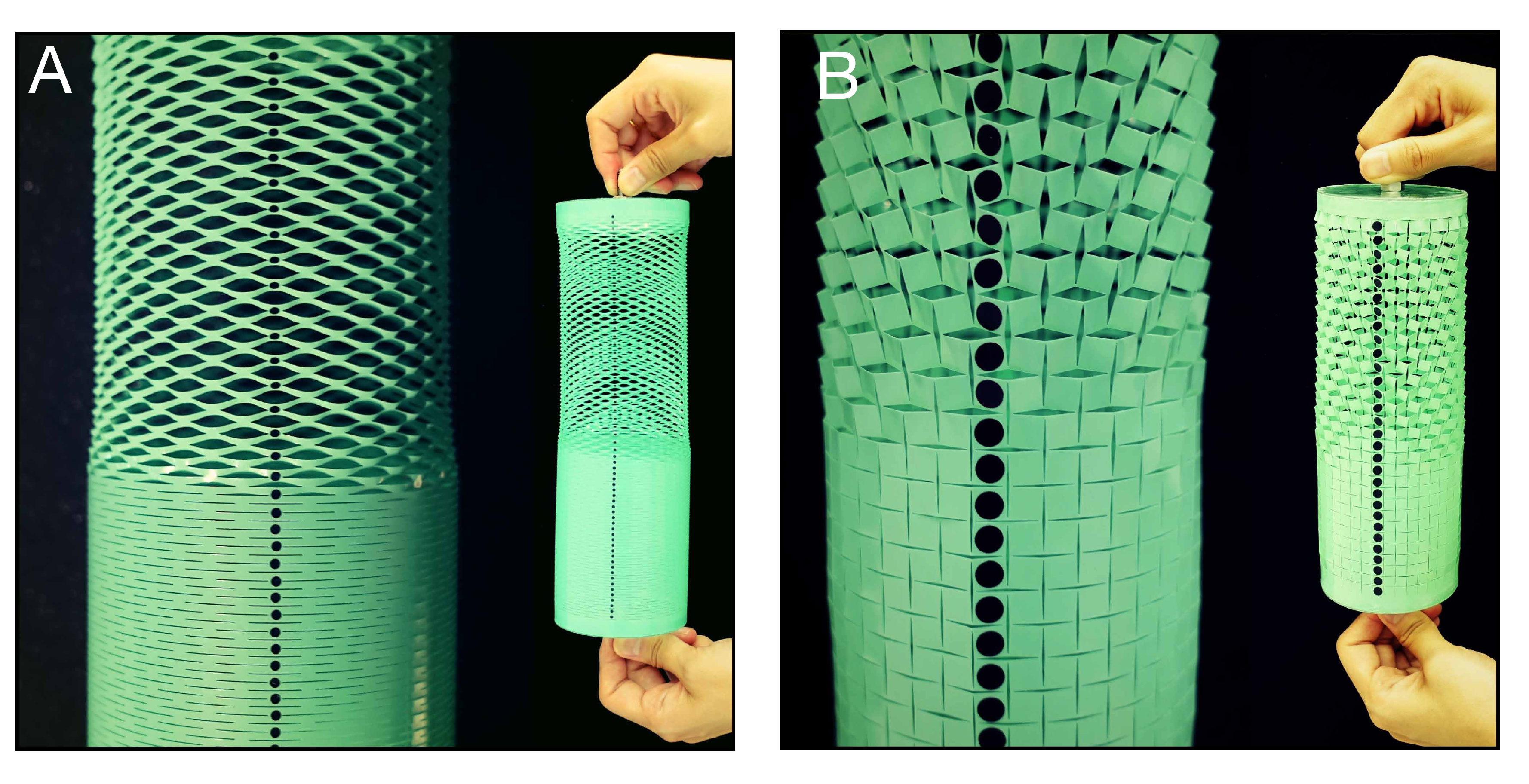}
\caption{(\textit{A}) Experimental images of a kirigami cylindrical shell with $n=20$ and linear cuts  characterized by $\delta/l=0.2$ (with $l=12$ mm) at $\bar\varepsilon\simeq0.2$.  (\textit{B}) Experimental  images of a kirigami cylindrical shell with $n=20$ and orthogonal cuts  characterized by $\delta/l=0.08$ (with $l=6$ mm) at $\bar\varepsilon\simeq0.1$. 
The thickness of kirigami shells is $t=76.2\:\mu$m.}
\label{Fig5}
\end{figure}

\section*{Effect of geometry}
While in Figs.~\ref{Fig4}\textit{B} we focus on a specific geometry, it is important to point out that our model can be used to efficiently characterize the  propagation front  as a function of both the curvature of the shell, the hinge size and the arrangement of the cuts.  In Fig.~\ref{Fig4}\textit{C} we focus on kirigami structures with triangular cuts and report the evolution of the normalized width of the propagation front, $W$ (which is defined as the width of  region in which the strain changes by $0.9 (\varepsilon_{p3}-\varepsilon_{p1})$ - see Fig.~\ref{Fig4}\textit{B}), as a function of $\delta/l$ and $n$.
First, the  results of our model indicate that as the curvature of the kirigami shell increases (i.e.~for larger $n$), $W$  monotonically increases, so that propagation of instability becomes less and less visible. In the limit of flat sheets all unit cells are characterized by monotonic stress-strain curves  and only homogeneous pop-up is possible (see white region in Fig.~\ref{Fig4}\textit{C}). Second, we find that by increasing $\delta/l$, at constant $n$, the propagation of instabilities becomes more accentuated as the width $W$ of the transition zone  monotonically decreases (see Fig.~\ref{Fig4}\textit{C}). It is also interesting to note that the width of the transition zone is inversely proportional to the  energy barrier $\mathcal{S}_1$ (see  Fig.~\ref{Fig4}\textit{D}). The largest values of $\mathcal{S}_1$ are observed for unit cells with  very large  $\delta/l$ and very small $n$. For such units  the peak in stress is followed by a sharp drop and the mechanical response is characterized by a discrete sequence of drops during propagation (see SI Appendix, Fig.~S15\textit{D}), each corresponding to the opening of one row of cuts (see Movie~S4).

Notably, using our model in combination with FE analysis conducted on  the unit cells phase diagrams similar to those shown in Figs.~\ref{Fig4}\textit{C} and \textit{D} can be constructed for any cut pattern (see SI Appendix, Fig.~S16). 
Such diagrams can then be used to identify  regions in the parameter space where propagation of instability is triggered. As examples, in Figs.~\ref{Fig5}\textit{A} and~\ref{Fig5}\textit{B} we report snapshots of cylindrical kirigami shells with a staggered array of linear cuts~\cite{Shyu2015, Blees2015, Isobe2016} and  an array of mutually orthogonal cuts~\cite{Cho2014, Rafsanjani2017}, respectively. Both images clearly show the coexistence of the popped and unpopped phases (see SI Appendix, Figs.~S5~and~S6, and Movies~S5-S7) and further indicate that the characteristics of the phase transition can be controlled by carefully selecting the geometry of the cuts as well as the curvature of the shell. The kirigami shell with the linear pattern is characterized by a sharp propagation front spanning across about one unit cell and a propagation stress $\sigma_p=177\:$kPa (see SI Appendix, Fig.~S18), whereas the orthogonal cuts lead to a wider front spreading across about four unit cells and $\sigma_p=320\:$kPa (see SI Appendix, Fig.~S19). 

\begin{figure}[!t]
\centering
\includegraphics[width=0.9\columnwidth]{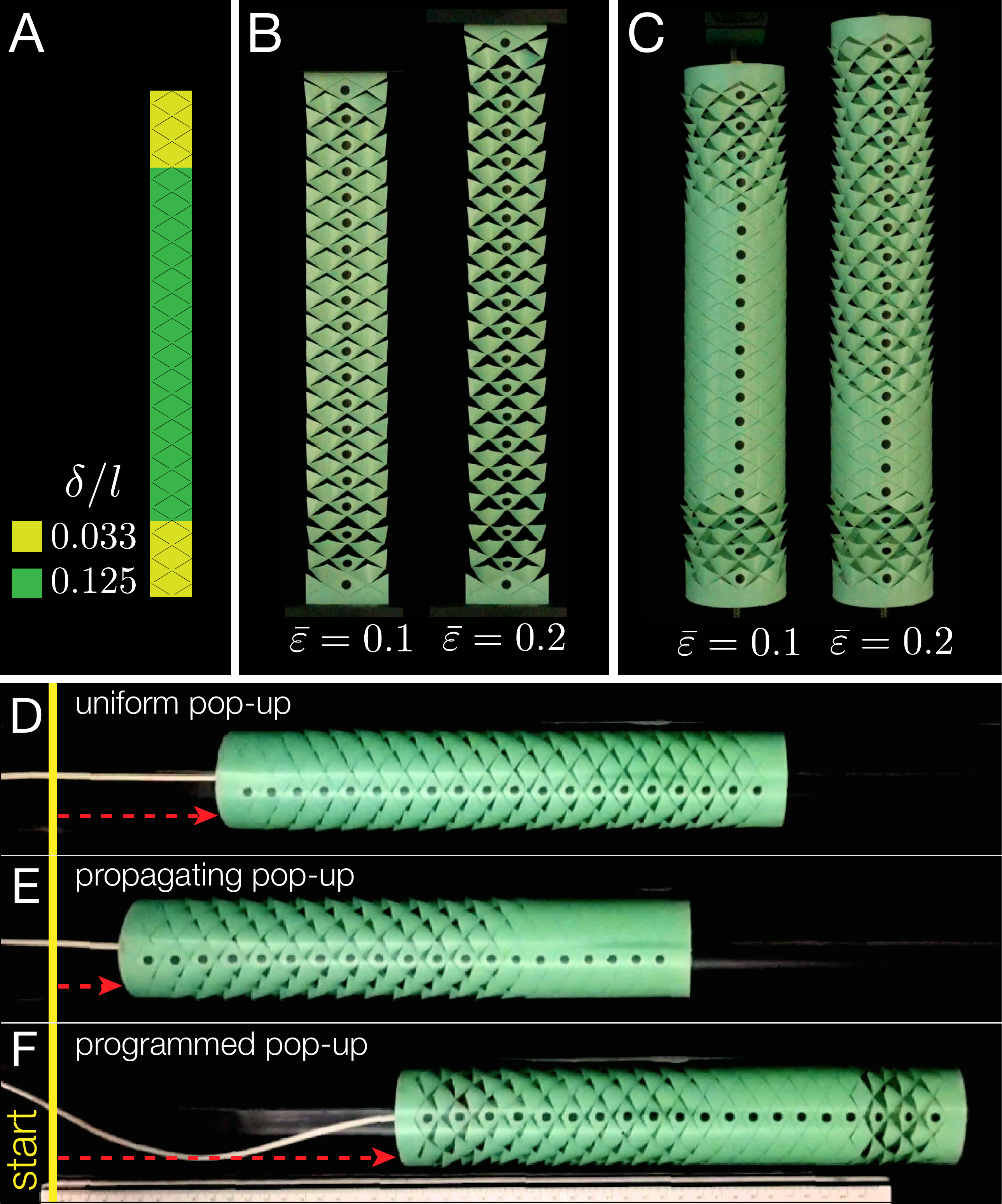}
\caption{(\textit{A}) Schematic of our patterned kirigami surface. (\textit{B})-(\textit{C}) Experimental images of the heterogenous (\textit{B}) kirigami sheet and (\textit{C}) kirigami shell at $\bar\varepsilon=0.1$ and 0.2. (\textit{D})-(\textit{F}) Snapshots of our kirigami-skinned crawlers with triangular cuts characterized by  (\textit{D}) $\delta/l=0.033$ (uniform pop-up), (\textit{E}) $\delta/l=0.125$ (propagating pop-up) and (\textit{F}) $\delta/l=0.125$ for the central units and  $\delta/l=0.033$ for first three units at two ends (programmed pop-up).}
\label{Fig6}
\end{figure}

Finally, we find that the coexistence of the buckled and unbuckled phases observed in our kirigami cylindrical shells  provides opportunities to realize surfaces with complex behavior that can be programmed to achieve a desired functionality. To demonstrate this, we consider a kirigami surface with $20$ rows of triangular cuts separated by hinges with  two different sizes. Specifically, we choose $\delta/l=0.033$ for three rows near the two ends and $\delta/l=0.125$ for the central rows (see Fig.~\ref{Fig6}\textit{A}). If such surface is planar, no clear signature of the two different $\delta/l$ is apparent upon stretching. Since the critical strains associated to the two considered $\delta/l$ are very close to each other ($\bar{\varepsilon}_{c}=4.4275\times10^{-4}$ and $3.510\times10^{-4}$ for $\delta/l=0.033$ and 0.125, respectively), all features pop-up almost simultaneously and tilt fairly uniformly (see Fig.~\ref{Fig6}\textit{B} and SI Appendix, Fig.~S9\textit{B}).  
By contrast, if we use the heterogeneous kirigami sheet to form a cylinder with $n=8$, upon stretching we find a clear sequence. First, the triangular features separated by ligaments with  $\delta/l=0.033$ pop up all together. Second, the pop-ups propagate in the central region with $\delta/l=0.125$ starting from the top (see Fig.~\ref{Fig6}\textit{C}, SI Appendix, Fig.~S9\textit{C} and Movie S8). Remarkably, this sequencing achieved by simply patterning the sheet with regions characterized by different ligament widths can be exploited to design a smart skin that significantly enhance the crawling efficiency of a linear actuator (Fig.~\ref{Fig6}\textit{D}-\textit{F}). 
While all three kirigami-skinned crawlers advance upon elongation and contraction of the actuator because of the    anisotropic friction induced by the pop-ups~\cite{Rafsanjani2018} (see SI Appendix, section 3), the programmed pop-up achieved in our patterned shell enhances the anchorage of the crawler to the substrate at two ends and significantly reduces the backslide (see SI Appendix, Fig.~S11\textit{B}). As a result, the patterned crawler (Fig.~\ref{Fig6}\textit{F}), which benefits from coexistence of popped and unpopped regions at desired locations, proceeds about twice faster than the crawlers with an homogeneous array of triangular cuts with either $\delta/l=0.033$ (Fig.~\ref{Fig6}\textit{D}) or $\delta/l=0.125$ (Fig.~\ref{Fig6}\textit{E}).


\section*{Discussion and conclusions}
\begin{figure}[!t]
\centering
\includegraphics[width=\columnwidth]{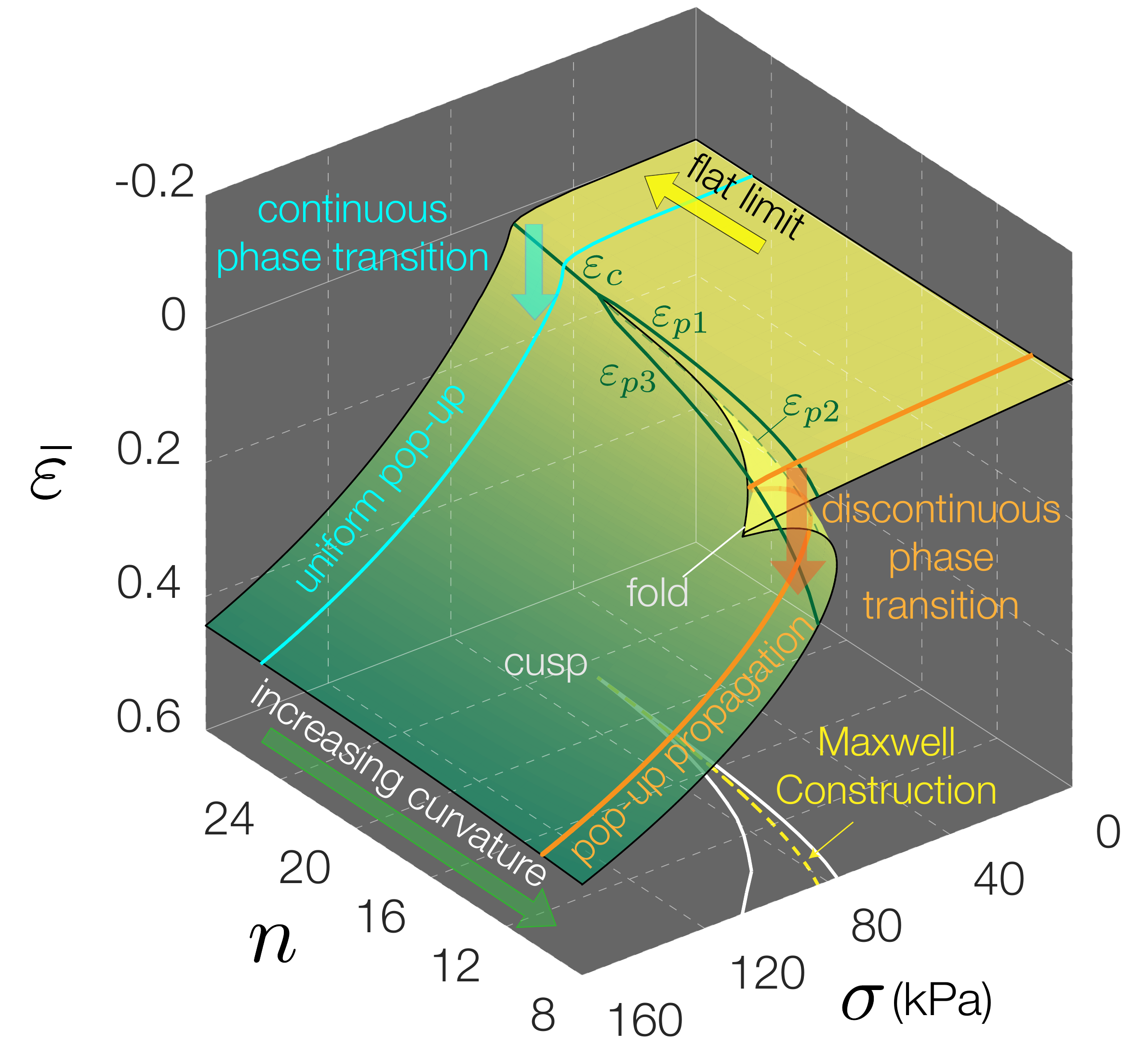}
\caption{Behavior surface for cylindrical kirigami shells with triangular cuts characterized by $l=12\:$mm, $\delta/l=0.0625$ and $t=76.2\:\mu$m. The continuous sequence of states of equilibrium experienced by the shell upon loading
corresponds to a curve on the behavior surface. The behavior surface is smooth, but its projection on $n$-$\sigma$ contains two
folds and a cusp. 
For shells with small curvature (i.e. for large $n$) the path does not reach the folds (see blue line) and uniform pop-ups occurs at a critical strain $\varepsilon_c$ associated with out-of-plane buckling of hinges. By increasing the curvature of the shell, the path eventually intersects  the folds (see red line).  Inside the cusp, popped and unpopped regions coexist and pop-ups can propagate with a deformation path that passes from a stable unpopped strain $\varepsilon_{p1}$ to another stable popped strain $\varepsilon_{p3}$ through an unstable intermediate strain $\varepsilon_{p2}$.  The projection of the $\varepsilon_{p1}$ (or $\varepsilon_{p2}$, $\varepsilon_{p3}$) curves on the $\sigma-n$ plane represents the loci of propagation stress given by Maxwell construction.}
\label{Fig7}
\end{figure}

To summarize, we have shown that in cylindrical kirigami shells the buckled and unbuckled phases can coexist, with the pop-up process initially starting near an end and then propagating along the cylinder at constant stress.  In contrast to flat kirigami sheets, which can only support continuous phase transitions, by introducing curvature the buckling-induced transformation exhibits discontinuity in the first derivative of the free energy, resulting in the coexistence of two phases~\cite{Jaeger1998}. 
This remarkable difference in behavior arises because the curvature transforms  the ligaments from straight columns that buckle to  bistable  arches that snap.
It should be also noted that such response is completely different from that of porous cylindrical shells which under compression exhibit uniform buckling-induced shape transformation~\cite{Matsumoto2012, Lazarus2015, Javid2016},  whereas it shares similarities with structures consisting of an array of beams resting on flexible supports, which have recently shown to exhibit a very rich response~\cite{Xu2015, Abdullah2016}. 
The behavior of our system can be further understood by looking at its  behavior surface (see Fig.~\ref{Fig7} for a triangular pattern with $\delta/l=0.125$). We find that by increasing the curvature of the shell (i.e.~by decreasing $n$)  a \textit{cusp catastrophe} emerges~\cite{Zeeman1976, Thompson1975, Lu2016}. The increase in curvature  causes a progressively larger divergence between the top and bottom faces of the fold, making the discontinuous phase transition more pronounced.
As such, the behavior surface of Fig.~\ref{Fig7} further confirms that the curvature is the essential ingredient to trigger propagation of pop-ups. However, we also find that the stresses introduced to bend the sheets into cylinders play an important role, as they increase the energy barrier and make the propagation more pronounced (see SI Appendix, Fig.~S14). Finally, we have shown that the characteristics of discontinuous phase transition  can be tuned by carefully selecting the geometry of the kirigami structure.
With such control on the phase transition in kirigami structures, we envision that these mechanical metamaterials could be used to design the next generation of responsive surfaces, as shown by the design of a smart skin that enhances the crawling efficiency of a linear actuator. 

\matmethods{Details of fabrication of kirigami shells are described in SI Appendix, section 1. 
The protocol for experiments and additional experimental data for kirigami shells with triangular, linear, and orthogonal cut patterns are provided in SI Appendix, section 2. Principles of Kirigami-skinned crawlers are presented in SI Appendix, section 3. Details of FE simulations and theoretical model are presented in SI Appendix, section 4.} 

\showmatmethods{} 

\acknow{We thank J. W. Hutchinson for fruitful discussions and Lisa Lee, Omer Gottesman, and Shmuel M. Rubinstein for technical support and access to their custom laser profilometer. 
A.R.~acknowledges support from Swiss National Science Foundation Grant P300P2-164648. 
L.J.~acknowledges support from National Natural Science Foundation of China Grants 11672202, 11602163, and 61727810.
K.B.~acknowledges support from NSF Grant DMR-1420570 and Army Research Office Grant W911NF-17-1-0147. 
The views and conclusions contained in this document are those of the authors and should not be interpreted as representing the official policies, either expressed or implied, of the Army Research Laboratory or the US Government. The US Government is authorized to reproduce and distribute reprints for government purposes notwithstanding any copyright notation herein.}

\showacknow{} 


\end{document}